\documentclass[12pt]{article}

\usepackage{amsfonts, amsmath, amssymb, bm, latexsym}
\usepackage{esint} 
\usepackage{xcolor}
\usepackage[hidelinks]{hyperref}
\usepackage{slashed} 

\newcommand{\beq}{\begin{equation}}
\newcommand{\eeq}{\end{equation}}
\newcommand{\bea}{\begin{eqnarray}}
\newcommand{\eea}{\end{eqnarray}}

\usepackage[lmargin=1.0in, rmargin=1.0in, tmargin=1.0in, bmargin=1.0in]{geometry}

\usepackage{setspace}
\begin{document}



\title{\bf Foldy--Wouthuysen Transformation of the Generalized Dirac Equation in Symmetric Teleparallel Gravity}
\author{Sabutay Ugur$^1$, Muzaffer Adak$^{1,*}$, Ali Bagci$^1$, Caglar Pala$^2$  \\
   {\small $^1$Computational and Gravitational Physics Laboratory, Department of Physics,} \\
   {\small Faculty of Sciences, Pamukkale University, Denizli, Türkiye} \\
  {\small $^2$Department of Physics,
Faculty of Arts and Sciences, Erciyes University, Kayseri, Türkiye} \\
      {\small {\it E-mail:} {yatubas@gmail.com, madak@pau.edu.tr, abagci@pau.edu.tr, caglar.pala@gmail.com}} \\
      {\small $^*$Corresponding author: madak@pau.edu.tr}}

  \vskip 1cm
\date{Revised September 2026}
\maketitle
\thispagestyle{empty}

\begin{abstract}
\noindent
We investigate the non-relativistic limit of the generalized Dirac equation in a weak, static, and spherically symmetric background of symmetric teleparallel gravity. The underlying generalized spinor connection incorporates the complete Clifford-algebra basis and introduces additional couplings to the non-metricity sector beyond those of the conventional Dirac theory. Working in the coincident gauge and adopting the weak-field Schwarzschild geometry in isotropic coordinates, we derive the corresponding generalized Dirac Hamiltonian and perform successive Foldy--Wouthuysen transformations up to order $1/m^2$, retaining terms to first order in the gravitational potential and its spatial derivatives. The resulting block-diagonal Hamiltonian contains not only the expected gravitational counterparts of the kinetic, spin--orbit, and Darwin interactions, but also additional operator structures generated by the generalized spinor connection. In particular, direct spin--gravity, anisotropic spin--momentum--gravity, and tidal spin--momentum couplings arise naturally from the generalized metric-affine interaction. We further perform an order-of-magnitude analysis for an electron in the Earth's weak gravitational field to justify the adopted truncation of the inverse-mass expansion. These results demonstrate that the generalized Dirac equation in a symmetric teleparallel background gives rise to new low-energy interaction channels involving the fermion spin, momentum, and spatial derivatives of the gravitational field. The resulting effective Hamiltonian provides a framework for exploring phenomenological constraints on the additional couplings entering the generalized spinor connection. \\
  


\end{abstract}

\section{Introduction}
\label{sec:introduction}

General Relativity (GR) has provided an extraordinarily successful classical description of gravitation and has been verified by a wide range of experimental and observational tests. Nevertheless, several fundamental problems remain unresolved, including the nature of dark matter and dark energy, the origin of the late-time accelerated expansion of the Universe, and the formulation of a consistent quantum theory of gravity \cite{RubinFord1970,Riess1998,RovelliVidotto2014}. These open questions have motivated extensive studies of modified and extended theories of gravity, including Chern--Simons models, non-minimal curvature-electromagnetic couplings, and scalar--tensor theories \cite{adak-dereli2012,sert-adak2019}. Parallel efforts have also been pursued within non-Riemannian formulations of gravity \cite{adak-dereli-senikoglu2019}. Among the various geometric frameworks proposed for this purpose, metric-affine geometry offers one of the most comprehensive settings by treating the metric and the affine connection as independent dynamical variables.

Within metric-affine geometry (MAG), spacetime may simultaneously possess curvature, torsion, and non-metricity, each representing an independent geometric characteristic of the affine structure \cite{hehl-mccrea1995,dereli-tucker1995}. These geometric features have also been incorporated into two-dimensional gravity through a Poisson--sigma model formulation \cite{adak-grumiller2007}. This richer geometry naturally enlarges the possible interactions between gravitation and matter. Fermionic fields are of particular interest because their dynamics depend explicitly on the spinor connection through the covariant derivative. Consequently, the Dirac equation provides a natural framework for investigating how non-Riemannian geometry may manifest itself in the dynamics of spin-$1/2$ particles \cite{ponomarev-obukhov1982,adak-sert2025}.

Motivated by this viewpoint, we recently formulated the most general covariant Dirac equation in a generic metric-affine spacetime containing curvature, torsion, and non-metricity \cite{adak-sert2025}. In that work, the spinor connection was generalized by incorporating all independent basis elements of the Clifford algebra, yielding the most general Clifford algebra-valued spinor connection compatible with MAG. The use of the full $16$-dimensional Clifford basis is a completeness statement: it spans the endomorphisms of the Dirac spinor space and prevents admissible covariant structures from being excluded at the outset. It does not imply that all Clifford grades are equally fundamental or that every coupling must be nonzero. Particular dynamics, symmetries, or experimental bounds may consistently select a constrained subspace of this general parametrization. The resulting Dirac equation contains several additional interaction terms governed by independent coupling parameters. While these terms arise naturally from the generalized geometric structure of spacetime, their physical significance has not yet been systematically investigated. Understanding their low-energy consequences is therefore essential for assessing the physical viability of the generalized Dirac theory.

As a first step toward exploring the physical implications of the generalized Dirac equation, we recently investigated neutrino propagation in a weak gravitational field within the framework of symmetric teleparallel gravity (STPG) formulated in the coincident gauge \cite{cetinkaya-adak2026}. In STPG, curvature and torsion vanish identically, while non-metricity remains as the sole nontrivial geometric field strength. This considerably simplifies the underlying geometry without eliminating the effects associated with the generalized spinor connection. Throughout our analysis, the coincident gauge is imposed in the coordinate frame, whereas all calculations are performed in the corresponding orthonormal frame. That study demonstrated that the additional coupling parameters can influence neutrino oscillations in weak gravitational fields, providing the first indication that the generalized Dirac equation may lead to observable effects beyond those predicted by the conventional minimally coupled theory. For discussions of STPG and the natural (coincident) gauge, see Refs.~\cite{adak2005,adak-sert2006,adak2006-tjp,adak-sert2013,adak2018,koivisto-jimenez2018,adak-pala2023sce,heisenberg2024}.

We emphasize that the gravitational sector used here is the symmetric teleparallel equivalent of GR, often called coincident GR, rather than a modified $f(Q)$ theory. It reproduces the same Schwarzschild metric and classical gravitational predictions as GR, while representing the gravitational field through non-metricity instead of curvature. This equivalent formulation is useful as a controlled test bed: it keeps the background metric familiar and isolates the consequences of the generalized fermion connection in a purely non-metricity sector.

The physical implications of the additional interactions predicted by our recently proposed generalized Dirac equation remain to be fully understood. Thus, a natural starting point is the existing body of work on the interaction between fermionic spin and gravitational fields, which provides the appropriate theoretical framework for interpreting the low-energy consequences of the generalized spinor couplings. The non-relativistic limit of the Dirac equation in gravitational fields has therefore been studied extensively over the past several decades. Early work by Hehl and Ni demonstrated how inertial and gravitational effects enter the Dirac equation and influence the dynamics of relativistic spin-$1/2$ particles \cite{hehl-ni1990}. Subsequently, Obukhov developed a covariant description of spin dynamics in external gravitational fields and clarified the corresponding Foldy--Wouthuysen (FW) representation \cite{obukhov2001,obukhov2002}. These studies were further extended by Silenko and Teryaev, who established the semiclassical limit of the Dirac equation in gravitational backgrounds and demonstrated the correspondence between quantum spin dynamics and the classical motion of spinning particles \cite{silenko-teryaev2005}. Later investigations generalized these analyses to rotating sources, strong gravitational fields, and arbitrary stationary spacetimes, leading to increasingly complete descriptions of spin precession, gravitational Stern--Gerlach forces, and the validity of the equivalence principle for relativistic fermions \cite{obukhov-silenko-teryaev2009,obukhov-silenko-teryaev2011,obukhov-silenko-teryaev2013}.

Despite these important developments, almost all existing FW analyses, including our earlier study in a non-Riemannian spacetime with torsion and non-metricity \cite{adak-dereli2003}, are based on the conventional covariant Dirac equation. In our previous work, the spinor connection was obtained through the standard Kosmann lift of the spacetime connection, and the resulting Hamiltonian reflected only the couplings arising from the conventional non-Riemannian geometry. By contrast, the recently proposed generalized metric-affine Dirac equation \cite{adak-sert2025} incorporates the complete Clifford-algebra basis in the spinor connection together with independent coupling parameters, thereby predicting a broader class of low-energy interactions. The physical consequences of these generalized couplings have not previously been investigated within the FW formalism. The present work fills this gap and may be regarded as the phenomenological continuation of our recent theoretical formulation.

The FW transformation provides a systematic procedure for separating the positive and negative-energy sectors of the Dirac Hamiltonian and deriving its non-relativistic limit \cite{foldy-wouthuysen1950,bjorken-drell1964}. Beyond its mathematical role as a block-diagonalization technique, the FW representation is particularly valuable because it expresses relativistic dynamics in terms of effective low-energy operators with clear physical interpretations. Kinetic, spin-dependent, relativistic, and external-field interactions emerge explicitly in the transformed Hamiltonian, making the FW representation especially suitable for identifying and interpreting the physical effects associated with the additional couplings introduced by the generalized spinor connection.

In the present work, we investigate the non-relativistic limit of the generalized Dirac equation \cite{adak-sert2025} in a weak, static, and spherically symmetric background of STPG formulated in the coincident gauge. Starting from the generalized covariant Dirac equation, we derive the corresponding Dirac Hamiltonian and perform successive FW transformations up to order $1/m^2$. We then carry out an order-of-magnitude analysis for an electron in the Earth's weak gravitational field to establish the hierarchy of the relevant energy scales and justify the adopted truncation of the inverse-mass expansion. Finally, we compare the resulting effective Hamiltonian with the familiar FW Hamiltonian of a Dirac particle interacting with an electromagnetic field, thereby identifying the gravitational counterparts of the standard kinetic, spin--orbit, and Darwin interactions, as well as the additional operator structures generated by the generalized spinor connection. In particular, we show that the generalized metric-affine couplings naturally give rise to direct spin--gravity, anisotropic spin--momentum--gravity, and tidal spin--momentum interactions that are absent in the conventional theory. The restricted background is deliberately chosen as a benchmark rather than as a claim of universality: it makes the new operator structures explicit before extensions to rotating, time-dependent, or fully metric-affine backgrounds are attempted. The results are therefore relevant both to quantum fields in non-Riemannian geometry and to precision searches for anomalous spin-dependent interactions.

\section{Symmetric Teleparallel Background Geometry}
\label{sec:background_stpg}

We consider the gravitational field generated by a static, spherically symmetric source of mass $M$. In particular, when the source is identified with the Earth, its orbital motion around the Sun and its rotation about its own axis are neglected. The gravitational background is therefore assumed to be described by the Schwarzschild geometry. Within the symmetric teleparallel equivalent of GR, the metric solution coincides with the Schwarzschild solution of GR, although the corresponding affine connection and non-metricity structures are different. In isotropic coordinates $x^\mu=(ct,x,y,z)$, the line element is written as
  \begin{align}
    ds^2 =  -f^2(r)c^2dt^2 + g^2(r) \left( dx^2+dy^2+dz^2 \right),
   \end{align}
where $ r=\sqrt{x^2+y^2+z^2} $, and the metric functions are
  \begin{align}
    f(r) &= \frac{1+\xi}{1-\xi}, &  g(r) &=  (1-\xi)^2, &  \xi  &=     -\frac{GM}{2rc^2}.
   \end{align}
An orthonormal coframe, $e^a$, adapted to this metric in terms of the coordinate coframe, $dx^\mu$, may be chosen as
   \begin{align}
    e^0 = fc\,dt \qquad \text{and} \qquad e^i = g\,dx^i. 
    \label{eq:on-coframe}
  \end{align}
Throughout this paper, the first half of the lowercase Latin alphabet, $a,b,\ldots,h=0,1,2,3$, is used for orthonormal spacetime indices, while the remaining lowercase Latin letters, $i,j,\ldots=1,2,3$, are reserved for spatial indices. The orthonormal coframe and the coordinate coframe are, in general, related through the tetrads $h^a{}_\alpha$ and their inverses $h^\alpha{}_a$ according to
  \begin{align}
    e^a = h^a{}_\mu dx^\mu \qquad \text{and} \qquad dx^\mu = h^\mu{}_a e^a.
   \end{align}
For the above coframe, they explicitly take the form
  \begin{align}
    h^a{}_\mu =
    \begin{pmatrix}
        f&0&0&0\\
        0&g&0&0\\
        0&0&g&0\\
        0&0&0&g
    \end{pmatrix}
     \qquad \text{and} \qquad
    h^\mu{}_a
    =
    \begin{pmatrix}
        1/f&0&0&0\\
        0&1/g&0&0\\
        0&0&1/g&0\\
        0&0&0&1/g
    \end{pmatrix}.
   \end{align}

We adopt the coincident gauge, in which the affine connection vanishes in the coordinate frame \cite{adak2006-tjp,koivisto-jimenez2018}. Upon transforming to the orthonormal frame, however, the corresponding connection $1$-forms become non-vanishing and are generated entirely by the change of frame,
  \begin{align}
    \omega^a{}_b = h^a{}_\mu \, dh^\mu{}_b.
  \end{align}
For the tetrads given above, this yields
\begin{align}
    \omega^a{}_b
    =
    \begin{pmatrix}
        -\dfrac{f'}{rfg}x_i e^i&0&0&0\\
        0&-\dfrac{g'}{rg^2}x_i e^i&0&0\\
        0&0&-\dfrac{g'}{rg^2}x_i e^i&0\\
        0&0&0&-\dfrac{g'}{rg^2}x_i e^i
    \end{pmatrix},
\end{align}
where a prime denotes differentiation with respect to $r$ and $ x_i e^i := xe^1+ye^2+ze^3$. The non-metricity $1$-forms are defined by the symmetric part of the full connection, $Q_{ab} := \omega_{(ab)} = \frac{1}{2} \left( \omega_{ab}+\omega_{ba} \right)$. The two trace $1$-forms of non-metricity relevant to the generalized Dirac equation are consequently
\begin{align}
    Q := \eta^{ab}Q_{ab} = A(r)x_i e^i \qquad \text{and} \qquad 
    P:= \left( \iota^aQ_{ab} \right)e^b = B(r)x_i e^i,
    \label{eq:traces-of-nonmetricity}
\end{align}
where
\begin{align}
    A(r) := -\frac{1}{rg} \left( \frac{f'}{f} + 3\frac{g'}{g}
    \right) \qquad \text{and} \qquad
    B(r) := -\frac{g'}{rg^2}.
\end{align}
On the other hand, the antisymmetric part of the connection vanishes identically in the chosen orthonormal frame, $ \omega_{[ab]} = \frac{1}{2} \left( \omega_{ab}-\omega_{ba} \right) = 0$. Thus, for the particular symmetric teleparallel background and frame adopted here, the geometric contributions entering the generalized Dirac equation are carried by the non-metricity sector, in particular through the trace $1$-forms $Q$ and $P$. This considerably simplifies the construction of the corresponding Dirac Hamiltonian and its subsequent FW transformation.

   \section{Dirac Hamiltonian} \label{sec:dirac_hamiltonian}

We now derive the Dirac Hamiltonian for a spin-$1/2$ particle of mass $m$ propagating in the symmetric teleparallel background introduced in the previous section. Starting from the generalized Dirac equation obtained in our previous work \cite{adak-sert2025} and imposing the symmetric teleparallel conditions $T^a=0$ and $R^a{}_b=0$, we write
   \begin{align}
    *\gamma\wedge D\psi+m\psi *1=0, \qquad D\psi=(d+\Omega)\psi, \label{eq:gen-dirac-eqn}
   \end{align}
where $\psi$ denotes the spinor field, $*$ is the Hodge map, $\gamma=\gamma_a e^a$ is the Clifford-algebra-valued $1$-form, and $*1$ is the volume form. The generalized spinor connection $1$-form is given by
  \begin{align}
  \Omega = \frac{1}{2}\sigma_{ab}\omega^{[ab]} + (a_1 I+a_2\gamma_5)Q
 + (a_3 I+a_4\gamma_5)P + (b_3\gamma_a+b_4\gamma_a\gamma_5)e^a . \label{eq:generalized-spinor-connection}
  \end{align}
Here $\gamma_a$ are the Dirac matrices satisfying $ \{\gamma_a,\gamma_b\}=2\eta_{ab}I$, while $ \gamma_5:=\gamma_0\gamma_1\gamma_2\gamma_3$ and $\sigma_{ab}:=\frac{1}{4}[\gamma_a,\gamma_b]$. The Clifford algebra $cl(1,3)$ is generated by the Dirac matrices $\gamma_a$ together with the identity and may be represented by the $16$-dimensional basis $ \left\{ I, \,
\gamma_a, \, \sigma_{ab}, \, \gamma_a\gamma_5, \, \gamma_5 \right\}$. The coupling parameters $a_1$, $a_2$, $a_3$, and $a_4$ are dimensionless, whereas $b_3$ and $b_4$ have dimensions of inverse length. Whenever a dimensionless parametrization is useful, we introduce $\bar b_3 := \ell_P b_3$ and $\bar b_4 := \ell_P b_4$, where $\ell_P = \sqrt{{\hbar G}/{c^3}}$ is the Planck length. Unless otherwise stated, we retain the dimensional quantities $b_3$ and $b_4$ throughout the formal derivation.

The six-parameter expression in Eq.~\eqref{eq:generalized-spinor-connection} is the STPG restriction of the general Clifford-valued connection constructed in Ref.~\cite{adak-sert2025}; it is not postulated as an independent gauge field with separate field equations. In that construction, all Clifford sectors allowed by covariance are first retained, and agreement between the Dirac equation obtained by direct minimal coupling and the equation obtained from the generalized Dirac Lagrangian imposes consistency conditions on the couplings. The coefficients that remain unfixed by this comparison should therefore be viewed, at the present stage, as phenomenological parameters. A more fundamental matter-sector dynamics or additional symmetry may further restrict them.

For the symmetric teleparallel background considered in Sec.~\ref{sec:background_stpg}, the antisymmetric part of the connection vanishes $\omega^{[ab]}=0$. Equation~\eqref{eq:gen-dirac-eqn} therefore reduces to
  \begin{align}
   *\gamma\wedge d\psi = *\gamma\wedge \Big[ -(a_1 I+a_2\gamma_5)Q - (a_3 I+a_4\gamma_5)P - (b_3 I-b_4\gamma_5)\gamma \Big] \psi - m\psi *1. \label{eq:dirac-stpg-intermediate}
  \end{align}
For the orthonormal coframe given in Eq.~\eqref{eq:on-coframe}, the exterior derivative of the spinor field can be expressed as
  \begin{align}
  d\psi = \frac{1}{fc} \frac{\partial\psi}{\partial t}e^0 + \frac{1}{g} \frac{\partial\psi}{\partial x^i}e^i. \label{eq:dpsi-orthonormal}
  \end{align}
In the intermediate steps below, the factor $c$ is temporarily suppressed for notational simplicity and restored in the final Hamiltonian. Using Eq.~\eqref{eq:traces-of-nonmetricity} together with the identities
 \begin{align}
   *e^a\wedge e^b &= -\eta^{ab}*1, &  \gamma_a\gamma^a &= 4I, &  \gamma_a\gamma_5\gamma^a &= -4\gamma_5,
 \end{align}
Eq.~\eqref{eq:dirac-stpg-intermediate} reduces to
  \begin{align}
   -\frac{1}{f}\gamma^0\partial_t\psi = \frac{1}{g} \gamma^i \partial_i \psi + (a_1A+a_3B)x_i\gamma^i \psi + (a_2A+a_4B) x_i\gamma^i \gamma_5\psi + 4b_3\psi + 4b_4\gamma_5\psi - m\psi, \label{eq:dirac-before-hamiltonian}
  \end{align}
where we have introduced the abbreviations $\partial_t\psi := 
{\partial\psi}/{\partial t}$ and $\partial_i\psi := {\partial\psi}/{\partial x^i}$. To cast the generalized Dirac equation into Hamiltonian form, we introduce the Bjorken--Drell matrices according to our conventions, $ \beta := i\gamma_0$ and $\alpha_i := \gamma_0\gamma_i$, together with the linear momentum operator in position space, $p_i=-i\hbar\partial_i$. Multiplying Eq.~\eqref{eq:dirac-before-hamiltonian} from the left by the appropriate factor containing $f\gamma^0$ and restoring $\hbar$ and $c$, we obtain the Schrödinger form $ i\hbar {\partial\psi}/{\partial t}=H\psi$, where the generalized Dirac Hamiltonian is
  \begin{align}
   H &= \frac{f}{g}c \,\alpha^i p_i - ic\hbar f(a_1A+a_3B)x_i\alpha^i \nonumber \\
     &\quad - ic\hbar f(a_2A+a_4B)x_i\alpha^i\gamma_5 - 4c\hbar b_3f\beta - 4c\hbar b_4f\beta\gamma_5 + fmc^2\beta . \label{eq:hamilton-stpg}
 \end{align}

Equation~\eqref{eq:hamilton-stpg} constitutes the relativistic Hamiltonian from which the FW transformation will be performed. Its structure already reveals the different origins of the individual contributions. The first term is the kinetic term modified by the gravitational background through the factor $f/g$. The combinations $ a_1A+a_3B$ and $a_2A+a_4B$ encode the couplings of the spinor field to the two independent traces of non-metricity, $Q$ and $P$. The former multiplies the ordinary $\alpha^i$ structure, whereas the latter involves the additional chiral matrix $\gamma_5$. The terms proportional to $b_3$ and $b_4$ originate from the additional Clifford-valued sector of the generalized spinor connection. They play qualitatively different roles. The $b_3$ contribution, $-4c\hbar b_3f\beta$, has a direct mass-like structure, whereas the $b_4$ contribution, $-4c\hbar b_4f\beta\gamma_5$, contains the chiral matrix $\gamma_5$ and therefore introduces a chirality-sensitive mass-like coupling. As will become apparent after the FW transformation, both parameters contribute to effective rest-energy shifts as well as to additional momentum- and spin-dependent interactions. Finally, the term $fmc^2\beta$ represents the conventional rest-mass contribution modified by the gravitational background.

For completeness, we now state the reality conditions and the inner product relevant to the Hermiticity of Eq.~\eqref{eq:hamilton-stpg}. On a static spatial hypersurface $\Sigma_t$, the natural Dirac inner product reduces to
\begin{align}
 \langle \phi,\psi\rangle_{\Sigma_t}
   = \int_{\Sigma_t} d^3x\,g^3(r)\,\phi^\dagger\psi .
 \label{eq:curved-inner-product}
\end{align}
Consequently, the adjoint of $p_i=-i\hbar\partial_i$ with respect to this product, denoted by $\ddagger$, is
\begin{align}
 p_i^{\ddagger}=p_i-i\hbar\,\partial_i\ln g^3 .
 \label{eq:weighted-momentum-adjoint}
\end{align}
This measure contribution is essential. Defining $F=f/g$ and $C_i=(a_1A+a_3B)x_i$, Hermiticity of the sum of the kinetic term and the term proportional to $C_i\alpha^i$ is equivalent, up to the usual boundary conditions, to
\begin{align}
 \partial_iF+F\partial_i\ln g^3=2f\,\operatorname{Re}(C_i).
 \label{eq:hermiticity-condition}
\end{align}
Using the definitions of $A(r)$ and $B(r)$, Eq.~\eqref{eq:hermiticity-condition} gives
\begin{align}
 a_1=-\frac{1}{2}+i\widetilde a_1,\qquad
 a_3= \frac{1}{2}+i\widetilde a_3,\qquad
 a_2,a_4,b_3,b_4\in\mathbb{R},\qquad
 \widetilde a_1,\widetilde a_3\in\mathbb{R}.
 \label{eq:coupling-reality-conditions}
\end{align}
These are precisely the direct--variational consistency conditions inherited from Ref.~\cite{adak-sert2025}. With Eq.~\eqref{eq:coupling-reality-conditions}, the remaining algebraic terms in Eq.~\eqref{eq:hamilton-stpg} are also Hermitian with respect to Eq.~\eqref{eq:curved-inner-product}. Notice in particular that $a_1$ and $a_3$ are not purely imaginary: their fixed real parts compensate the adjoint contribution from the position-dependent kinetic operator and the spatial measure.

The dimensional structure of the last two generalized couplings deserves particular attention. Since $b_3$ and $b_4$ have dimensions of inverse length, their characteristic energy scales are $E_{b_3} \sim \hbar c\,b_3$ and $E_{b_4} \sim \hbar c\,b_4$. In terms of the dimensionless parameters introduced above, these can equivalently be written as $  \hbar c\,b_3 = \bar b_3E_P$ and $ \hbar c\,b_4 = \bar b_4E_P$, where $E_P = {\hbar c}/{\ell_P}$ is the Planck energy. This distinction between the dimensional couplings $b_3,b_4$ and their dimensionless counterparts $\bar b_3,\bar b_4$ will be important when estimating the characteristic magnitudes of the additional interaction terms.

\section{Foldy--Wouthuysen Transformation}
\label{sec:FW-transformation}

The FW transformation provides a systematic procedure for obtaining the non-relativistic limit of the Dirac theory by separating the positive- and negative-energy sectors of the Hamiltonian. When the characteristic momentum of a massive fermion satisfies $|\vec{p}|\ll mc$, the four-component Dirac description can be reduced, order by order in inverse powers of the mass, to an effectively block-diagonal form \cite{foldy-wouthuysen1950,bjorken-drell1964}. In the following, we apply this procedure to the generalized Dirac Hamiltonian given by Eq.~(\ref{eq:hamilton-stpg}).

For the formal FW calculation, we temporarily adopt natural units $\hbar=c=1$. The fundamental constants will be restored whenever physical energy scales are discussed. We write the Hamiltonian in the standard form
 \begin{align}
    H = \beta m+\vartheta+\varepsilon, \label{eq:FW-general-H}
 \end{align}
where $\vartheta$ and $\varepsilon$ denote the odd and even operators, respectively, satisfying $\{\beta,\vartheta\}=0$ and $[\beta,\varepsilon]=0$. All adjoints and FW transformations below are understood with respect to the inner product in Eq.~\eqref{eq:curved-inner-product}. Under the conditions in Eq.~\eqref{eq:coupling-reality-conditions}, $\vartheta$ and $\varepsilon$ are self-adjoint and the successive FW transformations are unitary in this Hilbert-space structure.

For the weak gravitational field considered here, the Schwarzschild metric functions in isotropic coordinates can be expanded as
\begin{align}
    f &\simeq  1+V,  &  g &\simeq 1-V,  &  \frac{f}{g}  &\simeq  1+2V,  \label{eq:weak-field-metric-functions}
\end{align}
where the dimensionless Newtonian potential is $V(r) = -{GM}/{rc^2}$. Accordingly, $\vec{\nabla}V := -{\vec{g}}/{c^2}$. Throughout this section, all expressions are retained only to first order in the gravitational potential and its derivatives. Thus, terms of the form $V^2$, $V\partial_iV$, $(\partial_iV)(\partial_jV)$, and analogous higher-order combinations are systematically neglected.

The weak-field expansions of the geometrical functions introduced in Sec.~\ref{sec:background_stpg} are
\begin{align}
    \frac{f'}{f} &\simeq V',  &  \frac{g'}{g}  &\simeq  -V',  &  fA &\simeq \frac{2V'}{r},  &  fB  &\simeq \frac{V'}{r}.
\end{align}
Using $\partial_iV =  V'{x_i}/{r}$, the combinations that appear in the generalized Dirac Hamiltonian become
   \begin{subequations}
     \begin{align}
    (a_1fA+a_3fB)x_i &\simeq  (2a_1+a_3)\partial_iV, \\
    (a_2fA+a_4fB)x_i  &\simeq  (2a_2+a_4)\partial_iV.
     \end{align}
   \end{subequations}
It is, therefore, convenient to introduce the effective coupling combinations
   \begin{align}
    u := 2a_1+a_3 =-\frac{1}{2}+i\left(2\widetilde a_1+\widetilde a_3\right) \in\mathbb{C} \qquad \text{and} \qquad v &:= 2a_2+a_4 \in\mathbb{R}.  \label{eq:u-v-definitions}
   \end{align}
Thus, the $u$-dependent momentum and Laplacian terms appearing below must be read as a jointly self-adjoint operator combination; they need not be Hermitian term by term when written in an unsymmetrized form.
Now, according to our conventions,
   \begin{align}
    \alpha_i\gamma_5 &= \gamma_5\alpha_i := i\Sigma_i,  &   [\beta,\Sigma_i]  &= 0,   &   \{\beta,\beta\gamma_5\} &= 0.
   \end{align}
Hence, $\alpha_i$ and $\beta\gamma_5$ are odd, whereas $\beta$ and $\Sigma_i$ are even. Consequently, to first order in the weak gravitational field, the odd and even parts of the Hamiltonian in Eq.~\eqref{eq:FW-general-H} are
  \begin{subequations} \label{eq:stpg-even-odd-matrices1}
    \begin{align}
    \vartheta &\simeq  \left[ (1+2V)p_i - iu(\partial_iV) \right] \alpha^i - 4b_4(1+V) \beta \gamma_5, \label{eq:stpg-even-odd-matrices1-odd} \\
    \varepsilon &\simeq \left[ mV - 4b_3(1+V) \right]\beta + v (\partial_iV) \Sigma^i. \label{eq:stpg-even-odd-matrices1-even}
    \end{align}
  \end{subequations}

Since the background is static, no explicit time derivative of the FW transformation operator appears. The first FW transformation is defined by
  \begin{align}
    H' = e^{iS}He^{-iS}, \qquad  S = -\frac{i}{2m}\beta\vartheta. \label{eq:first-FW-generator}
  \end{align}
Using the Baker–Campbell–Hausdorff expansion,
  \begin{align}
    H' &= H + i[S,H] -\frac{1}{2}[S,[S,H]] - \frac{i}{6}[S,[S,[S,H]]] + \frac{1}{24}[S,[S,[S,[S,H]]]] + \cdots, \label{eq:stpg-H-prime1}
  \end{align}
and retaining terms through order $1/m^2$, we obtain
  \begin{align}
    H' \simeq  \beta m+\varepsilon'+\vartheta',
  \end{align}
where
   \begin{subequations}
     \begin{align}
    \varepsilon' &= \varepsilon + \frac{1}{2m}\beta\vartheta^2 - \frac{1}{8m^2} [\vartheta,[\vartheta,\varepsilon]], \\
    \vartheta' &=  \frac{1}{2m}\beta[\vartheta,\varepsilon] - \frac{1}{3m^2}\vartheta^3.
    \end{align}
   \end{subequations}
The first transformation removes the leading odd contribution $\vartheta$, but it generates the higher-order odd operator $\vartheta'$.

A second FW transformation is, therefore, performed according to
  \begin{align}
    H'' = e^{iS'}H'e^{-iS'},  \qquad  S' = -\frac{i}{2m}\beta\vartheta'. \label{eq:second-FW-generator}
  \end{align}
Retaining only terms that can contribute through order $1/m^2$, we find
  \begin{align} 
    H''  \simeq  \beta m+\varepsilon''+\vartheta'', 
  \end{align}
with
   \begin{subequations}
     \begin{align}
    \varepsilon'' &= \varepsilon + \frac{1}{2m}\beta\vartheta^2 - \frac{1}{8m^2}
    [\vartheta,[\vartheta,\varepsilon]], \\
    \vartheta''  &= \frac{1}{4m^2}  [[\vartheta,\varepsilon],\varepsilon].
    \end{align}
  \end{subequations}
The remaining odd contribution is of order $1/m^2$ and can be removed by a third FW transformation,
   \begin{align}
    H''' = e^{iS''}H''e^{-iS''}, \qquad  S'' = -\frac{i}{2m}\beta\vartheta''. \label{eq:third-FW-generator}
   \end{align}
Thus, up to order $1/m^2$, the block-diagonal Hamiltonian becomes
  \begin{align}
    H'''_{\rm STPG} \simeq \beta m + \varepsilon + \frac{1}{2m} \beta \vartheta^2 - \frac{1}{8m^2} [\vartheta , [\vartheta , \varepsilon ] ]. \label{eq:H-triple-prime}
  \end{align}

Equation~\eqref{eq:H-triple-prime} is the basic FW expression to be evaluated for the present STPG background. Substituting the explicit operators in Eqs.~\eqref{eq:stpg-even-odd-matrices1-odd} and \eqref{eq:stpg-even-odd-matrices1-even}, expanding the operator products and commutators, and retaining terms only through order $1/m^2$ and first order in the gravitational field, we obtain
   \begin{align}
   H'''_{\rm STPG}  \simeq{}& \Bigg\{ m - 4b_3 + \frac{8b_4^2}{m} + \frac{32b_3b_4^2}{m^2} + \left[ m - 4b_3 + \frac{16b_4^2}{m} - \frac{8b_4^2(m-12b_3)}{m^2} \right]V \nonumber \\
   & \quad + \left[ \frac{1}{2m} + \frac{2b_3}{m^2} \right] \vec{p}^{\,2} + \left[ \frac{3}{2m} + \frac{10b_3}{m^2} \right] V \vec{p}^{\,2} - i \left[ \frac{1+2u}{2m} + \frac{(6+4u)b_3}{m^2} \right] \vec{\nabla} V \cdot \vec{p} \nonumber \\
   &\quad + \left[ -\frac{u}{2m} + \frac{m - 4b_3 - 16b_3u + 8b_4v}{8m^2} \right] \nabla^2V + \left[ \frac{3}{4m} + \frac{5b_3+2b_4v}{m^2} \right] \vec{\Sigma} \cdot (\vec{\nabla}V \times \vec{p}) \Bigg\}\beta \nonumber\\
   &+ \left[ v + \frac{b_4}{m} + \frac{12b_3b_4}{m^2} \right] \vec{\Sigma} \cdot \vec{\nabla}V - \frac{v}{2m^2} \left[ (\vec{\Sigma} \cdot \vec{\nabla}V) \vec{p}^{\,2} - (\vec{\nabla}V \cdot \vec{p}) (\vec{\Sigma} \cdot \vec{p}) \right] \nonumber\\
   &+ \frac{iv}{4m^2} \Sigma^j(\partial_i\partial_jV)p_i - \frac{iv}{4m^2} (\nabla^2V) \vec{\Sigma} \cdot \vec{p} + \frac{v}{8m^2} \vec{\Sigma} \cdot \vec{\nabla}(\nabla^2V). \label{eq:H-triple-prime-compact}
   \end{align} 
Equation~\eqref{eq:H-triple-prime-compact} is the central result of the FW analysis. It explicitly displays the non-relativistic operator structures induced by the generalized spinor connection in the weak, static, and spherically symmetric STPG background. In addition to the familiar gravitational modifications of the kinetic and spin–orbit terms, the generalized connection generates additional spin–gravity, momentum–gravity, spin–momentum–gravity, and higher-derivative interactions governed by the effective coupling combinations $u$ and $v$ and by the dimensional parameters $b_3$ and $b_4$. The physical interpretation of these terms and their comparison with the standard electromagnetic FW Hamiltonian are discussed in the Discussion section.

\section{Validity of the $1/m^2$ Truncation}
\label{sec:validity-truncation}

Before discussing the physical implications of the effective Hamiltonian, it is useful to estimate the characteristic energy scales relevant to the physical system under consideration. These estimates provide a quantitative justification for truncating the FW expansion at order $1/m^2$. As a representative example, we consider an electron bound in a hydrogen atom located near the Earth's surface. The magnitude of the dimensionless gravitational potential at the Earth's surface is approximately $|V| = {GM_\oplus}/{R_\oplus c^2} \sim 10^{-9}$. The electron rest energy and the characteristic momentum scale in the hydrogen ground state are, respectively, $mc^2 \sim 10^{-13}\ {\rm J}$ and $pc \sim \alpha mc^2 \sim 10^{-16}\text{--}10^{-15}\ {\rm J}$, where $\alpha$ is the fine-structure constant. The corresponding gravitational potential-energy scale is $|mc^2V| \sim 10^{-22}\ {\rm J}$. The quantities $a_1,\ldots,a_4$, and hence the effective combinations $u$ and $v$, are dimensionless. By contrast, $b_3$ and $b_4$ have dimensions of inverse length. Their associated characteristic energy scales were given in the last paragraph of Sec. \ref{sec:dirac_hamiltonian}. 

For the purpose of examining the convergence of the FW expansion, we assume that the presently unconstrained generalized couplings do not generate contributions parametrically larger than the characteristic weak-gravity scale $|mc^2V|$. This should be regarded as an order-of-magnitude working assumption rather than as a phenomenological bound derived from experiment. Under this assumption, the characteristic hierarchy of energy scales may be represented schematically as
  \begin{align}
    mc^2 &\sim 10^{-13}\ {\rm J},   &  |\vartheta|  &\sim  10^{-16}\text{--}10^{-15}\ {\rm J},  &  |\varepsilon|  &\sim 10^{-22}\ {\rm J},
  \end{align}
where the dominant contribution to the odd operator $\vartheta$ is provided by the ordinary momentum term. Consequently, ${|\vartheta|}/{mc^2} \sim 10^{-3}\text{--}10^{-2}$, which provides the principal small parameter controlling the successive FW transformations.

The terms generated at successive stages of the FW expansion contain different combinations of $\vartheta$, $\varepsilon$, and their commutators, so their magnitudes cannot in general be inferred from a single universal suppression factor. Nevertheless, for the representative scales adopted above, the contributions retained through order $1/m^2$ may reach the approximate scale of $10^{-19}\ {\rm J}$, whereas representative contributions first entering at the next order are expected to be smaller, with characteristic magnitudes around $10^{-21}\ {\rm J}$. These estimates indicate a hierarchy of roughly two orders of magnitude between the relevant retained and neglected corrections for the physical regime considered here.

The truncation at order $1/m^2$ should therefore be understood as a controlled effective approximation designed to identify the leading non-relativistic operator structures generated by the generalized spinor connection. It does not imply that all higher-order contributions are physically irrelevant. In particular, terms beyond the present truncation may have to be retained in a dedicated analysis of fine-structure effects or in comparisons with sufficiently high-precision spectroscopic measurements.

\section{Discussion}
\label{sec:discussions}

To elucidate the physical content of the block-diagonal Hamiltonian given by Eq.~\eqref{eq:H-triple-prime-compact}, it is useful to compare its operator structure with the familiar FW Hamiltonian of a non-relativistic electron interacting with an electromagnetic field. Up to the corresponding order in the inverse-mass expansion, the electromagnetic Hamiltonian may be written schematically as \cite{bjorken-drell1964}
  \begin{align}
   H'''_{\rm em} ={}& \beta \left[ m + \frac{(\vec p-e\vec A)^2}{2m} - \frac{p^4}{8m^3} \right] + e\Phi - \frac{e}{2m}\beta\vec\Sigma\cdot\vec B \nonumber \\
   &- \frac{ie}{8m^2} \vec\Sigma \cdot (\vec\nabla \times \vec E) - \frac{e}{4m^2} \vec\Sigma \cdot (\vec E \times \vec p) - \frac{e}{8m^2} \vec\nabla \cdot \vec E . \label{eq:FW-electromagnetic-comparison}
  \end{align}
This Hamiltonian contains the familiar kinetic, Zeeman, spin-orbit, and Darwin structures. The STPG Hamiltonian derived in the present work exhibits several formally analogous operator structures, together with additional terms generated by the generalized spinor connection and its coupling to non-metricity. The comparison is therefore useful at the level of operator structure, although the physical origins of the electromagnetic and gravitational interactions are different.

One of the most recognizable terms in our result is $\vec\Sigma \cdot \left( \vec\nabla V \times \vec p \right)$, which is structurally analogous to the electromagnetic spin-orbit interaction $\vec\Sigma \cdot \left( \vec E \times \vec p \right)$. In the present case, however, its coefficient is not determined solely by the particle mass. From Eq.~\eqref{eq:H-triple-prime-compact}, the corresponding coefficient also contains the generalized coupling parameters $b_3$, $b_4$, and $v$. The generalized spinor connection may therefore modify the effective strength of the gravitational spin-orbit interaction. Since the expectation value of this operator depends on the orbital and spin quantum numbers of the state under consideration, such a modification could, in principle, produce state-dependent energy shifts. This suggests that sufficiently precise spectroscopic measurements may provide a possible route for constraining particular combinations of the generalized coupling parameters.

A qualitatively different structure is the direct spin-gravity coupling
  \begin{align}
    \left[ v + \frac{b_4}{m} + \frac{12b_3b_4}{m^2} \right] \vec\Sigma \cdot \vec\nabla V. \label{eq:direct-spin-gravity}
  \end{align}
The closest electromagnetic analogy is the Zeeman interaction $\vec\Sigma\cdot\vec B$, although the present term couples the spin directly to the spatial gradient of the gravitational potential rather than to a magnetic field. Restoring SI units, it is useful to define the dimensionless combination
\begin{align}
 \mathcal{C}_{\rm sg}:=v+\frac{\hbar b_4}{mc}
 +\frac{12\hbar^2b_3b_4}{m^2c^2}.
 \label{eq:Csg-definition}
\end{align}
The corresponding contribution to the Hamiltonian is then
\begin{align}
 H_{\rm sg}=\hbar c\,\mathcal{C}_{\rm sg}\,
 \vec\Sigma\cdot\vec\nabla V
 =-\frac{\hbar}{c}\,\mathcal{C}_{\rm sg}\,
 \vec\Sigma\cdot\vec g,
 \label{eq:spin-gravity-SI}
\end{align}
where $\vec\nabla V=-\vec g/c^2$. This parity-odd and time-reversal-odd operator has the same spin--gravity structure used in searches for anomalous couplings with comagnetometers \cite{terrano-romalis2022,zhang2023}. For spin eigenstates parallel and antiparallel to $\vec g$, Eq.~\eqref{eq:spin-gravity-SI} produces a splitting $|\Delta E|=2\hbar g|\mathcal{C}_{\rm sg}|/c$. A recent $^{129}$Xe and $^{131}$Xe comagnetometer search constrained an anomalous neutron-spin coupling to terrestrial gravity to below $5.3\times10^{-22}\,\mathrm{eV}$ at $95\%$ confidence level \cite{zhang2023}. Equation~\eqref{eq:spin-gravity-SI} provides a direct translation between such an energy sensitivity and the combination $\mathcal{C}_{\rm sg}$, once the fermion species to which the generalized coefficients refer is specified. We do not impose species universality here; therefore, the neutron result is quoted as an experimentally relevant scale rather than as a model-independent numerical bound on the electron coefficients used in Sec.~\ref{sec:validity-truncation}.

The terms proportional to $\nabla^2V$ are structurally analogous to the electromagnetic Darwin contribution $ -\frac{e}{8m^2} \vec\nabla \cdot\vec E$. Since the Newtonian potential satisfies the Poisson equation, $\nabla^2V$ is directly related to the local matter density generating the gravitational field. The corresponding terms in the STPG Hamiltonian may therefore be interpreted as gravitational Darwin-like contributions that probe the local source distribution rather than only the value or gradient of the gravitational potential.

An even higher-derivative structure appears through $\vec\Sigma \cdot \vec\nabla(\nabla^2V)$. This operator has no direct counterpart in the standard electromagnetic Hamiltonian displayed in Eq.~\eqref{eq:FW-electromagnetic-comparison}. Through the Poisson equation, it probes spatial variations of the source density and may therefore be interpreted as a spin-dependent response to inhomogeneities in the matter distribution.

Among the additional momentum-dependent structures generated by the generalized spinor connection are
  \begin{align}
    (\vec\Sigma \cdot \vec\nabla V)\vec p^{\,2} - (\vec\nabla V\cdot\vec p) (\vec\Sigma \cdot \vec p) \qquad \text{and} \qquad \Sigma^j (\partial_i\partial_jV)p_i.  \label{eq:hessian-spin-momentum}
  \end{align}
The first expression simultaneously couples the spin, momentum, and gravitational field. If the gradient of the potential is regarded locally as defining a preferred spatial direction, this operator distinguishes different components of the spin and momentum relative to that direction. It therefore represents an anisotropic spin-momentum-gravity interaction rather than a simple scalar gravitational correction. Such a structure can, in principle, distinguish quantum states with different angular-momentum or spin configurations even when their unperturbed energies coincide. Its coefficient is proportional to $v$, and measurements sensitive to this operator would therefore constrain a specific combination of the generalized spinor couplings. The second structure in Eq.~\eqref{eq:hessian-spin-momentum} contains the Hessian of the gravitational potential, $\mathcal{E}_{ij} := \partial_i \partial_jV$, which characterizes the spatial variation of the local gravitational field. It is therefore natural to interpret $\Sigma^j\mathcal{E}_{ij}p_i$ as a tidal spin-momentum coupling. Unlike an interaction depending only on the local value of $\vec\nabla V$, this term is sensitive to the inhomogeneity of the gravitational field and thus probes spatial variations that cannot, in general, be eliminated throughout a finite region by transforming to a freely falling frame.

The Hamiltonian also contains contributions proportional to $-i\,\vec\nabla V\cdot\vec p$. These terms arise naturally from the action of the momentum operator on the position-dependent geometrical quantities entering the Dirac Hamiltonian and from the corresponding operator products and commutators generated during the FW transformations. They should therefore not be regarded as independent potential-energy contributions, but rather as part of the complete momentum-dependent response of the fermion to the inhomogeneous gravitational background. In particular, they accompany terms involving $V\vec p^{\,2}$ and $\nabla^2V$, reflecting the common origin of these structures in the spatial dependence of the background geometry.

Another noteworthy consequence of the generalized spinor connection is that the parameters $b_3$ and $b_4$ generate non-vanishing contributions even in the formal limit $V\rightarrow0$. In natural units, the field-independent part of the positive-energy sector contains contributions of the form $-4b_3$, ${8b_4^2}/{m}$, and ${32b_3b_4^2}/{m^2}$. At first sight, these contributions may be interpreted as corrections to the effective rest energy of the fermion. However, $b_3$ and $b_4$ also enter the coefficients of the gravitational, kinetic, spin-orbit, and direct spin-gravity operators. Their physical effects, therefore, cannot, in general, be absorbed into a single redefinition of the particle mass. Instead, the same parameters control several distinct low-energy operator structures, making it possible, at least in principle, to constrain them through complementary physical observables.

The $b_3$ and $b_4$ interactions are not asserted to have been observed. They are allowed terms in the complete covariant parametrization and are retained so that their consequences can be confronted with experiment. The conventional sector is recovered by setting the additional coefficients to their standard values, in particular $b_3=b_4=0$; null results from spectroscopy, spin-precession measurements, and comagnetometers can then be used to restrict departures from this limit. The absence of an observed chirality-sensitive interaction is therefore a phenomenological constraint on the generalized theory rather than a reason to remove the corresponding operator before the comparison is made.

The choice of $cl(1,3)$ is fixed by the Lorentzian signature of the spacetime considered in this work and does not assign equal physical status to all Clifford grades. Formally replacing $\gamma_2\rightarrow i\gamma_2$ changes the real Clifford algebra to $cl(2,2)$ and hence changes the spacetime signature, the adjoint structure, and the causal interpretation of the Dirac theory. Although the complexified algebras are closely related, a $cl(2,2)$ construction is therefore a distinct geometric theory rather than a bivector-dominant subsector of the present Lorentzian problem. It would nevertheless be worthwhile to study, in future work, constrained ansatzes within $cl(1,3)$ in which particular Clifford sectors are selected by an additional symmetry or dynamical principle, as well as genuinely different-signature formulations when their physical setting is specified.

More broadly, the present analysis shows that STPG provides a particularly transparent setting in which the low-energy consequences of a generalized metric-affine coupling to fermions can be isolated and studied systematically. The resulting FW Hamiltonian demonstrates that non-metricity, when coupled through the generalized spinor connection, can generate qualitatively new spin-, momentum-, and tidal-dependent interaction channels beyond those arising from the conventional Dirac theory. Although the present investigation has been restricted to a weak, static, and spherically symmetric background, the formalism can be extended to more general metric-affine geometries in which curvature, torsion, and non-metricity coexist. Such extensions may help clarify which low-energy fermionic effects are characteristic of particular non-Riemannian sectors of spacetime geometry. The effective Hamiltonian derived here therefore provides a useful starting point for future phenomenological studies and for constraining the additional coupling parameters of the generalized Dirac equation.

\bigskip 
\noindent 
{\bf Acknowledgements.} This study was supported by the Scientific and Technological Research Council of Türkiye (TÜBİTAK) under Grant No.~124F325. The authors also thank the anonymous referees for their careful reading of the manuscript and for their constructive comments and suggestions, which helped improve the clarity and quality of this work.

  \bigskip 
 \noindent
{\bf Data Availability Statement.} Data sharing is not applicable to this article, as no datasets were generated or analyzed during the current theoretical study.

 \bigskip
 \noindent
{\bf Conflict of Interest.} The authors declare that they have no conflict of interest.

\bigskip
 \noindent
\textbf{Declaration of generative AI.}
The authors used ChatGPT (OpenAI; GPT-5.5 and GPT-5.6 Sol) to assist with language editing, improving readability and academic style, reorganizing parts of the presentation, supporting literature searches, and checking algebraic derivations involved in the Foldy--Wouthuysen calculations. All scientific arguments, equations, interpretations, and conclusions were critically reviewed by the authors, who take full responsibility for the accuracy and content of the manuscript.


\end{document}